\documentclass[aps,pre,twocolumn,10pt]{revtex4-1}
\usepackage{bm}
\usepackage{graphicx}
\usepackage{amsmath}
\usepackage{amssymb}
\usepackage{amscd}
\usepackage{epstopdf}
\usepackage{verbatim}
\usepackage[usenames]{color}
\usepackage{array}
\usepackage{xspace}
\usepackage{xstring}
\usepackage[normalem]{ulem}
\usepackage{wasysym}
\usepackage{bbold}
\usepackage{nicefrac}
\usepackage{layouts}

\newcommand{\ie}{\textit{i.e.}\xspace}
\newcommand{\eg}{\textit{e.g.}\xspace}

\newcommand{\B}{{\cal B}}

\newcommand{\Bb}{{\partial{\B}}}
\newcommand{\db}{\mathrm{d}}
\newcommand{\D}{n}
\newcommand{\fa}{F}        
\newcommand{\fu}{u}  
\newcommand{\fm}{f}        
\newcommand{\U}{U}      
\newcommand{\Ue}{\widetilde{U}} 
\newcommand{\ind}{I}  
\newcommand{\h}{H}  
\newcommand{\rhob}{\overline{\rho}}  
\newcommand{\kb}{{k_\text{B}}}

\definecolor{Blue}{rgb}{0,0,0.8}
\definecolor{BlueB}{rgb}{0,0,0.5}
\definecolor{Red}{rgb}{0.8,0,0}
\definecolor{RedB}{rgb}{0.6,0,0}
\definecolor{Green}{rgb}{0,0.5,0}
\definecolor{GreenB}{rgb}{0,0.8,0}
\definecolor{Purple}{rgb}{1,0,1}

\renewcommand{\vec}[1]{{\bf #1}}

\newcommand{\vecg}[1]{{\bm #1}}

\newcommand{\ddt}[1]{\dot{#1}}

\begin{document}

\title{Equilibrium mappings in polar-isotropic confined active particles}
\author{Yaouen Fily, Aparna Baskaran, Michael F. Hagan}
\affiliation{Martin Fisher School of Physics, Brandeis University, Waltham, MA 02453, USA}


\begin{abstract}
Despite their fundamentally non-equilibrium nature, the individual and collective behavior of active systems with polar propulsion and isotropic interactions (\emph{polar-isotropic active systems}) are remarkably well captured by equilibrium mapping techniques.
Here we examine two signatures of equilibrium systems -- the existence of a local free energy function and the independence of the coarse-grained behavior on the details of the microscopic dynamics -- in polar-isotropic active particles confined by hard walls of arbitrary geometry at the one-particle level.
We find that boundaries that possess concave regions make the density profile strongly dynamics-dependent and give it a nonlocal dependence on the geometry of the confining box.
This in turn constrains the scope of equilibrium mapping techniques in polar-isotropic active systems.
\end{abstract}

\maketitle



\section{Introduction} 
Active matter is a class of dissipative driven systems in which the driving forces are controlled by the microscale dynamics of the constituent particles~\cite{Marchetti2013}. Unlike an equilibrium system, which relaxes toward a unique free energy minimum that does not depend on the details of the dynamics, the steady-state properties of active systems are inextricably connected to their dynamics.
It is therefore remarkable that equilibrium concepts have been highly successful at explaining the behavior of the simplest form of active matter, self-propelled particles with isotropic interparticle interactions.
This is best illustrated by the phenomenon known as \emph{motility induced phase separation} (MIPS), in which self-propelled particles endowed with only repulsive interparticle interactions exhibit liquid-gas coexistence~\cite{Marchetti2016,Cates2015,Bialke2015}. While the equilibrium liquid-gas transition is driven by attractive interparticle interactions, MIPS arises because particle self-propulsion speeds effectively decrease with increasing local particle density. Despite this very different and non-equilibrium microscopic origin, MIPS exhibits all of the hallmarks of the equilibrium liquid-gas phase transition, and can be largely understood in terms of a local free energy function. Moreover,
the existence and phenomenology of MIPS is largely independent of the details of the particle dynamics, including the three commonly studied classes of self-propelled particles described below.

Despite the successes of equilibrium-like concepts, the intrinsically non-equilibrium nature of self-propelled particles suggests that equilibrium mappings must break down in some situations. In particular, the presence of an external confining potential (due to a boundary) has been shown to dramatically alter the behavior of self-propelled particles, in manners entirely unlike the effects of boundaries in an equilibrium liquid-gas system \cite{Galajda2007,Wan2008,Tailleur2009}. In this article we explore the applicability of equilibrium mappings for self-propelled particles confined within arbitrary boundary geometries.  We identify classes of boundaries in which equilibrium mappings break down, and associated qualitative changes in system behaviors.

Self-propelled particles with isotropic interparticle interactions correspond to the \emph{polar-isotropic} class of active particles (polar activity and isotropic interactions). Such particles exhibit persistent random motions, moving ballistically at short times and diffusively at long times. Three theoretical models for self-propelled particles have been extensively studied: (\textit{i}) In the active Brownian particle (ABP) model, which is motivated by autophoretic colloids, the propulsion speed is constant but the direction of motion diffuses due to thermal agitation.  (\textit{ii}) In the run-and-tumble particle (RTP) model, motivated by swimming bacteria such as \emph{E. coli}, particles move at constant speed and direction during `runs', broken by sudden turns (`tumbles').  (\textit{iii}) The active Ornstein-Uhlenbeck particle (AOUP) model describes passive particles pushed around by a bath of active particles, which behave like self-propelled particles driven by an active force which resembles a Gaussian colored noise~\cite{Maggi2014}. All three processes yield the same two-point time correlator of the active force, whose decay time $\tau$ sets the crossover from ballistic to diffusive motion, and respectively represents the inverse of the angular diffusion constant (ABP),  the average time between tumbles (RTP), and the memory time of the colored noise (AOUP). The decay time thus provides a mapping between the three models that can be used to probe the dependence of steady-state properties on the details of the stochastic dynamics.

The macroscale behaviors of the three models are consistent in bulk systems. For example, MIPS has been observed in all three models, and has been shown to have the same coarse grained description for ABP and RTP~\cite{Solon2015b}. However, in the presence of an external potential, this mapping is uncertain even at the one-particle level. Neither the conditions for existence of a local free energy nor the conditions under which the coarse-grained dynamics is independent of the details of the stochastic dynamics are known.
Maggi et al.'s work on AOUP suggests the existence of such a local free energy in a wide class of external potentials~\cite{Maggi2015}. Our work on persistent ($\tau\rightarrow\infty$) ABP confined within hard walls, on the other hand, shows the density profile is nonlocal when the walls are not convex~\cite{Fily2015}. Finally, Solon et al. found differences between the density profiles of ABP and RTP in uniform gravity fields and harmonic traps~\cite{Solon2015b}.

In this paper, we compare the analytical predictions of the AOUP theory of Ref.~\cite{Maggi2015} and those of the ABP theory of Refs.~\cite{Fily2014a,Fily2015,Fily2016} with each other and with numerical simulations of ABP, AOUP, and RTP in two and three dimensions for noninteracting particles confined within hard walls with various geometries.
When the walls are convex, we find perfect agreement between all available analytical and numerical results in both the quasi-thermal limit ($\tau\rightarrow0$) and the persistent limit ($\tau\rightarrow\infty$), regardless of the wall's geometry (as long as it is convex). In the intermediate $\tau$ regime, the decay of the bulk density with $\tau$ is model-dependent: it is exponential for ABP and algebraic for AOUP and RTP. The density profile on the wall, on the other hand, is largely model-independent and well captured by the AOUP theory, which implies the existence of an effective local free energy.
Conversely, for non-convex walls the density profile on the wall in the persistent regime becomes nonlocal and strongly model-dependent, thus inconsistent with a one-particle equilbrium mapping. This implies a breakdown of the AOUP theory, which we relate to the conditions of application of the unified colored noise approximation that it relies on.
In constrast, the ABP theory accomodates nonconvexity in the persistent limit, at least in two dimensions. It successfully predicts~\cite{Fily2015} the density profiles of ABP and AOUP, which are identical in the persistent limit, but not that of RTP, which is qualitatively different from the other two.
These results identify limits on the conditions for existence of a local free energy and a coarse-grained dynamics that is independent of model details for confined self-propelled particles, thus on the scope and applicability of equilibrium mapping techniques. 

The paper is organized as follows.
In section~\ref{model} we introduce a general polar-isotropic model for active particles of which ABP, RTP, and AOUP are three variants.
In section~\ref{convex_abp} we recall the theory of a very persistent ABP confined within hard convex walls~\cite{Fily2014a,Fily2016}, and apply it to generalized polar-isotropic active particles.
In section~\ref{convex_aoup} we extend the theory of an AOUP in a smooth external potential~\cite{Maggi2015} to the case of hard walls and compare it with the ABP theory.
In section~\ref{convex_simulations} we present Brownian dynamics simulations of ABP, AOUP, and RTP in elliptic and ellipsoidal boxes, and analyze the density profile on the boundary and the fraction of particles in the bulk.
In section \ref{nonconvex} we shift our attention to nonconvex containers. 
In section~\ref{nonconvex_abp} we review the theory for a very persistent ABP confined within hard nonconvex walls~\cite{Fily2015}, and discuss the conditions of its application to generalized polar-isotropic active particles.
In section~\ref{nonconvex_simulations} we present Brownian dynamics simulations of ABP, AOUP, and RTP in a 2D nonconvex box and analyze the density profile on the boundary.
In section~\ref{nonconvex_aoup} we show that the AOUP theory breaks down in nonconvex boxes.
Finally in section~\ref{conclusion} we discuss implications of these results for the concept of equilibrium mappings in polar-isotropic active systems.


\section{Model}
\label{model}

We consider an overdamped self-propelled particle in $\D=2$ or $3$ dimensions whose position $\vec{r}$ obeys the equation of motion
\begin{align}
\ddt{\vec{r}} = \mu \left( - \vecg{\nabla} \U + \vec{\fa} \right)
\label{eq:eom_r}
\end{align}
where $\mu$ is the mobility, $\U$ is the external potential, and the active force $\vec{\fa}$ obeys
\begin{align}
\langle \fa_i(t)\rangle = 0 
\, , \quad
\langle \fa_i(t)\fa_j(t')\rangle = \frac1\D \fm^2 \delta_{ij} e^{-|t-t'|/\tau}
\label{eq:fa_cor}
\end{align}
where $i$, $j$ are coordinate index, $\D$ is the dimension, $\fm$ is the root mean squared driving force, and $\tau$ is the persistence time.

Eq.~\eqref{eq:fa_cor} defines an entire class of polar-isotropic active models that share two key properties.
First, they have the same free mean squared displacement: if $\U=0$, then
$\langle[\vec{r}(t)-\vec{r}(0)]^2\rangle = 2\D D [t + \tau(1 - e^{-t/\tau})]$
where $D=\tau(\mu\fm)^2/\D$.
Thus, a free particle moves ballistically with typical speed $v=\mu\fm$ (typical for AOUP; ABP and RTP move at exactly $v$) at short times ($t\ll\tau$) and diffusively with diffusion constant $D$ at long times ($t\gg\tau$).
The typical distance travelled during a ballistic run, or persistence length, is $\ell=\tau\mu\fm=\sqrt{\D\tau D}$.
Second, they have the same quasi-thermal limit.
When $\tau$ is much smaller than every other time scale in the problem (or, equivalently, when $\ell$ is smaller than every length scale), the active force can be approximated as a white Gaussian noise and the system behaves like a thermal system with effective temperature $\kb T=D/\mu$.
The three most common self-propulsion models found in the literature all obey Eq.~\eqref{eq:fa_cor}. 
In the ABP model, inspired by autophoretic colloids, $\vec{\fa}$ has constant magnitude $\fm$ and its direction $\vec{u}=\vec{\fa}/|\vec{\fa}|$ undergoes rotational diffusion:
\begin{align}
\ddt{\vec{\fu}} = \frac{1}{\sqrt{(\D-1)\tau}} 
\left(1-\vec{\fu}\otimes\vec{\fu}\right)\vecg{\xi}
\label{eq:eom_u}
\end{align}
where $\vecg{\xi}$ is a white Gaussian noise with zero mean and correlations $\langle\xi_i(t)\xi_j(t')\rangle=2\delta_{ij}\delta(t-t')$
and $\left(1-\vec{\fu}\otimes\vec{\fu}\right)$ keeps $\vec{\fu}$ normalized.
In the RTP model, inspired by swimming bacteria such as \emph{E. coli}, $\vec{\fa}$ only changes through random instantaneous tumbles events that occur with a constant probability per unit time $\tau^{-1}$ and completely randomize $\vec{u}$ while leaving $|\vec{\fa}|$ unchanged.
In the AOUP model, which can be seen as a Gaussian approximation of either the ABP or the RTP model, $\vec{\fa}$ arises from the Ornstein-Uhlenbeck process
\begin{align}
\ddt{\vec{\fa}}
= - \frac{\vec{\fa}}{\tau} + \frac{f}{\sqrt{\D\tau}} \vecg{\xi}
\label{eq:eom_fa}
\end{align}

\textit{Confinement.} We consider a polar-isotropic particle that is confined to a region of space $\B$ (the box) by the elastic potential $\U(\vec{r})=\lambda \db(\vec{r})^2/(2\mu)$ where $\lambda$ is the wall's stiffness and $\db(\vec{r})=\min_{\vec{b}\in\B} |\vec{r}-\vec{b}|$ is the distance to the box.
Consistent with the requirements of the polar-isotropic model, this potential does not couple to the particle's orientation.
Both our ABP theory and our Brownian dynamics simulations assume perfectly hard walls ($\lambda\rightarrow\infty$), in which case the wall simply cancels the component of the velocity that is normal to the wall whenever the velocity points toward the wall. In the numerics, we simulate Eq.~\eqref{eq:eom_r} with $\phi=0$ but project any particle found outside the box back onto the boundary at the end of every time step.


\section{Convex walls}
\label{convex}

\subsection{Persistent hard-wall theory}
\label{convex_abp}

\begin{figure}
\centering
\includegraphics[width=\linewidth]{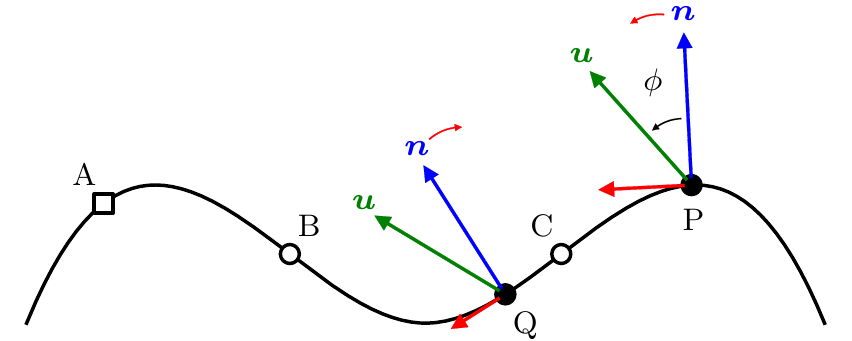}%
\caption{
Dynamics along the boundary. 
$\vec{n}$ is the local outward-pointing normal to the boundary. $\vec{u}$ is the particle's orientation. $B$ and $C$ are inflexion points. 
In a convex region (see $P$), the gliding motion rotates $\vec{n}$ toward $\vec{u}$ (red arrows), slowing down the particle.
In a concave region (see $Q$), the gliding motion rotates $\vec{n}$ away from $\vec{u}$, accelerating the particle.
In the absence of noise, a particle entering the concave region at $C$ with $\vec{u}\approx\vec{n}_\textsc{c}$ only comes to a rest at $A$ which verifies $\vec{n}_\textsc{a}=\vec{n}_\textsc{c}$.
}
\label{fig:wall_sketch}
\end{figure}

Let us first consider an infinitely persistent ($\tau\rightarrow\infty$) particle, \ie, one whose active force $\vec{\fa}$ is constant, confined within hard convex walls.
To achieve mechanical equilibrium, the particle must face the wall exactly, \ie, the local normal to the wall $\vec{n}$ must match the particle's orientation $\vec{u}$ (see Fig.~\ref{fig:wall_sketch}).
In a convex box, there is exactly one location where $\vec{n}=\vec{u}$ and, since the particle's motion along the boundary always rotates $\vec{n}$ toward $\vec{u}$ (see point $P$ in Fig.~\ref{fig:wall_sketch}), the equilibrium is always stable. We refer to it as the particle's \emph{target location}.

The effect of noise is to change $\vec{u}$, which in turn shifts the target location.
If the noise is weak, the target location moves slowly and the particle remains close to it ($\vec{n}\approx\vec{u}$)~\footnote{{This statement is to be understood in a statistical sense: $\lim_{\tau\rightarrow\infty}\langle\vec{n}-\vec{u}\rangle_\xi=0$.  For RTP, in particular, tumbles move the target location by a finite distance in zero time, making it impossible for the particle to be close to the target at all times.
However, the time it takes the particle to come back to the vicinity of the target after a tumble is independent of the noise, whereas the time between two tumbles goes to infinity in the persistent limit, therefore the fraction of the time the particle spends far from its target location vanishes in the persistent limit}}.
If the noise is also isotropic, \ie, if the steady-state distribution of $\vec{u}$ is uniform, so is that of $\vec{n}$, and the population of a patch of boundary is proportional to the number of normals it offers. Mathematically, the density on the boundary is equal to the Jacobian determinant of the function that maps a point on the boundary to its normal vector: in 2D, the curvature; in 3D, the Gaussian curvature.
In summary, the density profile of a single particle ($\int_{\vec{r}} \rho(\vec{r})=1$) in the weak noise limit is
\begin{align}
\begin{split}
& \rho(\vec{r}) = \int_{\vec{b}\in\Bb} \delta(\vec{r}-\vec{b})\ \rhob(\vec{b})
\, , \
\rhob(\vec{b}) = \frac{1}{S_n} \prod_{a=1}^{\D-1} \kappa_a(\vec{b})
\end{split}
\label{eq:persistent_density}
\end{align}
where $\Bb$ is the boundary, $\delta$ is the Dirac delta function, $\rhob$ is the density per unit length (in 2D) or area (in 3D) at the wall, $S_\D$ is the surface area of the $\D$-dimensional unit sphere, and $\kappa_a$ are the principal curvatures of the boundary.

Importantly, Eq.~\eqref{eq:persistent_density} is compatible with an equilibrium mapping:
it does not depend on the details of the noise as long as it is isotropic (which is true for ABP, RTP, and AOUP), and it takes the form $\rho(\vec{r})\propto e^{-\Ue(\vec{r})}$ with
\begin{align}
\Ue(\vec{r}) = -\log\left[\ind_\Bb(\vec{r})\prod_a\kappa_a(\vec{r})\right]
\label{eq:free_energy2}
\end{align}
where $\ind_\Bb(\vec{r})=1$ if $\vec{r}\in\Bb$ and $0$ otherwise is the indicator function of the wall, \ie, it minimizes the local effective free energy
\begin{align}
& {\cal F}[\rho] = \int_\vec{r} \Big[ \rho(\vec{r})\Ue(\vec{r}) -\rho(\vec{r})\big(\log\rho(\vec{r}) - 1\big) \Big]
\label{eq:free_energy1}
\end{align}
corresponding to an ideal gas in an external potential $\Ue$ with $\kb T=1$.


\subsection{AOUP theory}
\label{convex_aoup}

We now turn to the case of an AOUP, for which Maggi et al. derived a density profile $\rho$ in the presence of an external potential $\U$ that is exact in both the quasi-thermal limit ($\tau\rightarrow0$) and the persistent limit ($\tau\rightarrow\infty$) and approximate in between~\cite{Maggi2015}:
\begin{align}
\begin{split}
\rho(\vec{r}) & = \frac{\Omega(\vec{r})}{\int_{\vec{r}'} \Omega(\vec{r}')}
\\ 
\Omega(\vec{r}) & =
\det\big( \delta_{ij} + \tau\mu \partial_{ij} \U \big) \,
\exp\left[ -\frac{\mu\U}{D}-\frac{\tau\big|\mu\vecg{\nabla}\U \big|^2}{2D}  \right]
\end{split}
\label{eq:aoup_density_general}
\end{align}
where latin indices denote cartesian coordinates, $\det$ is the determinant, and $(\delta_{ij} + \tau \mu\partial_{ij} \U)$ should be positive definite.
It, too, is compatible with the effective ideal gas free energy \eqref{eq:free_energy1} with $\Ue=-\log\Omega$, as are all the density profiles we derive from it in the remainder of this section.

In theory, both Eq.~\eqref{eq:aoup_density_general} and Eq.~\eqref{eq:persistent_density} are exact in the $\tau\rightarrow\infty$ limit, and we expect them to agree with each other. However, the latter assumes hard walls whereas the former is formulated in terms of a smooth potential.
We circumvent this issue by evaluating Eq.~\eqref{eq:aoup_density_general} for the elastic wall potential introduced in section~\ref{model}, then taking the stiff limit ($\lambda\rightarrow\infty$) to obtain a hard wall prediction.

\begin{figure}
\centering
\includegraphics[width=0.9\linewidth]{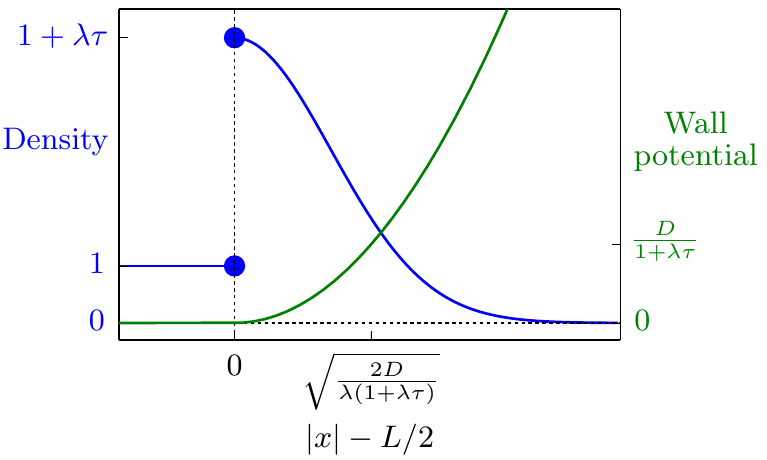}
\caption{Density profile (in blue) and wall potential (in green) for a 1D ideal gas of AOUP with unit bulk density near an elastic wall located at $x=L/2$ as predicted by Eq.~\eqref{eq:aoup_density_elastic_1D} for $\lambda\tau=4$.
}
\label{fig:wall_1d}
\end{figure}

Let us start with a box of size $L$ in $\D=1$ dimension.
Here $\U(x) = \frac12 \lambda\, (|x|-L/2)^2 \, \h(|x|-L/2)$ where $\h$ is the Heaviside function, thus
\begin{widetext}
\begin{align}
\Omega(x)
  = \left[ 1 + \lambda\tau \h\left(|x|-\frac{L}{2}\right) \right] \,
    \exp\left[ -\frac{\lambda(1+\lambda\tau)}{2D}\left(|x|-\frac{L}{2}\right)^2 \right]
  \xrightarrow[\lambda\rightarrow\infty]{}
  H\left(\frac{L}{2}-|x|\right) + \sqrt{\frac{\pi}{2}}\ \ell\ \delta\left(|x|-\frac{L}{2}\right)
\label{eq:aoup_density_elastic_1D}
\end{align}
\end{widetext}
The finite $\lambda$ case is shown in Fig.~\ref{fig:wall_1d}. In the stiff limit the density diverges at the wall, which hosts a finite population despite being reduced to a point.
In $\D=2$ and $3$ dimensions, the distance $\db$ used to define the confining potential remains constant as one moves parallel to the boundary and grows with slope $1$ as one moves perpendicular and away from the boundary, \ie, $\forall \vec{r}\notin\B$, $\vecg{\nabla} \db(\vec{r})=\vec{n}[\vec{b}(\vec{r})]$ where $\vec{b}$ is the point of the boundary closest to $\vec{r}$ and $\vec{n}$ is the normal to the boundary at $\vec{b}$~\cite{Gilbard1998}. It follows that, beyond the boundary,
$\partial_{ij}\db = (\delta_{ik} - n_i n_k) \partial_k n_j \equiv L_{ij}$
and $\partial_{ij} \U = \lambda ( n_i n_j + \db L_{ij} )$.
Here $L_{ij}$ is the shape operator at $\vec{b}$, whose eigenvalues along the principal directions of the boundary are the principal curvatures $\kappa_a$, completed by a zero eigenvalue along the normal $\vec{n}$.
Thus $\Omega(\vec{r})=1$ in the bulk and 
\begin{align}
\Omega(\vec{r})
& = \exp\left[ -\frac{1+\lambda\tau}{2D}\lambda \db^2 \right]
    \det\big[ \delta_{ij} + \lambda(n_i n_j + \db L_{ij}) \big]
\\
& = \exp\left[ -\frac{1+\lambda\tau}{2D}\lambda \db^2 \right]
    (1+\lambda\tau) \, \prod_a (1+\lambda\tau \db\kappa_a)
\label{eq:aoup_density_elastic}
\end{align}
beyond the boundary. As in the 1D case, in the $\lambda\rightarrow\infty$ limit
$\Omega$ yields a $\delta$ function at each point of the boundary whose weight is obtained by integrating over $\db$, leading to
\begin{align}
\begin{split}
& \Omega =
\ind_\B + \int_{\vec{b}\in\Bb} \delta(\vec{r}-\vec{b})\ \rhob_n(\vec{b})
\\
& \rhob_2 = \frac{\sqrt{\pi}}{2} \ell + \frac{\kappa}{2} \ell^2
,\
  \rhob_3 = \sqrt{\frac{\pi}{6}} \ell + \frac16 H \ell^2 + \sqrt{\frac{\pi}{54}} K \ell^3
\label{eq:aoup_density_stiff}
\end{split}
\end{align}
where $\kappa$ is the scalar curvature of the boundary in 2D, and $H=(\kappa_1+\kappa_2)/2$ and $K=\kappa_1 \kappa_2$ are the mean and Gaussian curvatures in 3D.
Like Eq.~\eqref{eq:persistent_density}, Eq.~\eqref{eq:aoup_density_stiff} only depends on the local geometry of the wall. On the other hand, it retains the finite $\ell$ corrections that make Eq.~\eqref{eq:aoup_density_general} exact in both the quasi-thermal and the persistent limit. For $\ell=0$, the integral in $\Omega$ vanishes and only the uniform bulk term survives, as expected for an equilibrium ideal gas. Conversely, for $\ell\rightarrow\infty$ the bulk term becomes negligible and the dominant boundary terms, of order $\ell^\D$, lead back to Eq.~\eqref{eq:persistent_density}.


\subsection{Simulations}
\label{convex_simulations}

\begin{figure}
\centering
\includegraphics[width=0.99\linewidth]{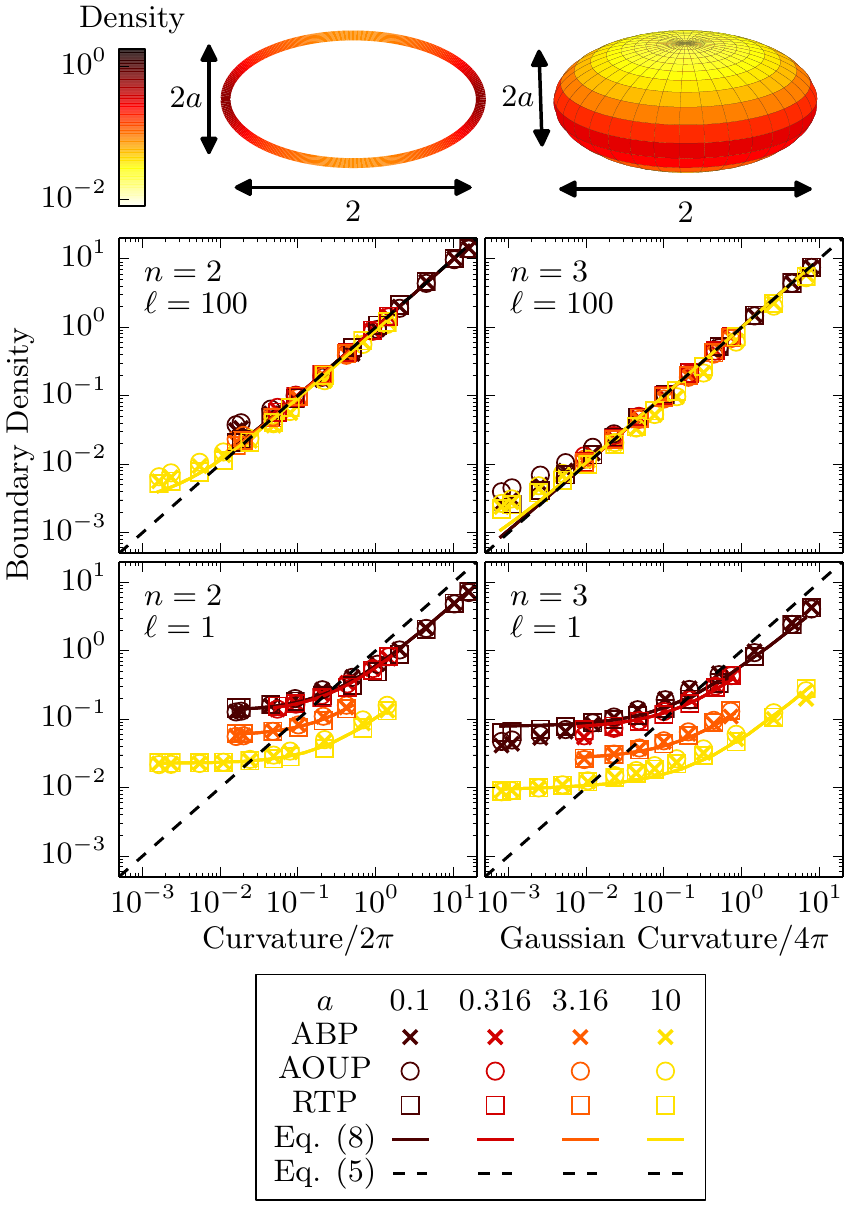}
\caption{
Density profile on the boundary of elliptic (left column) and spheroidal (right column) boxes.
The ellipses have semi-axes $(1,a)$. The spheroids have semi-axes $(1,1,a)$.
Each is shown at the top for $a=0.5$, with the persistent limit prediction [Eq.~\eqref{eq:persistent_density}] shown as a color map.
The four panels below show the simulated density of three types of self-propelled particles (ABP, RTP, AOUP) with persistence length $\ell=100$ (top row) and $\ell=1$ (bottom row) for various values of $a$.
In $\D=2$ dimensions (left column) we show the density per unit length of the boundary as a function of the local curvature.
In $\D=3$ dimensions (right column) we show the density per unit area of the boundary as a function of the local Gaussian curvature.
All three active models agree with each other and with the AOUP theory (Eq.~\eqref{eq:aoup_density_stiff}, solid lines) in all explored persistence regimes.
In the persistent regime (large $\ell$), the AOUP theory also agrees with the hard-wall persistent theory (Eq.~\eqref{eq:persistent_density}, dashed lines).
}
\label{fig:radius-density}
\end{figure}

\begin{figure}
\centering
\includegraphics[width=\linewidth]{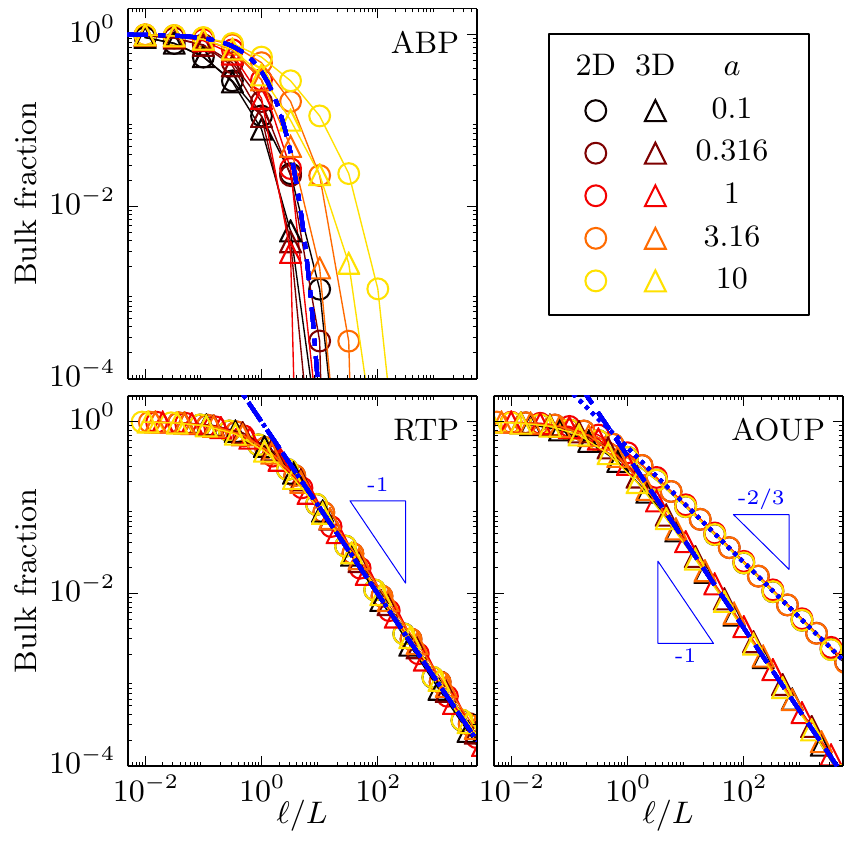}
\caption{
Fraction of particles in the bulk in the samples of Fig.~\ref{fig:radius-density} as a function of the persistence length $\ell$ divided by a characteristic confinement length scale $L$ (see text). The decay is roughly exponential in ABP (the dash-dotted line is $e^{-\ell/L}$) and algebraic in RTP and AOUP with exponent $-1$ (dashed lines) or $-2/3$ (dotted line).
}
\label{fig:bulk_fraction}
\end{figure}

The relationship between the density on the wall and the local curvature is illustrated in Fig.~\ref{fig:radius-density}, which compares Eq.~\eqref{eq:persistent_density} (dashed lines) and Eq.~\eqref{eq:aoup_density_stiff} (solid lines) with Brownian dynamics simulations of ABP (crosses), AOUP (circles), and RTP (squares) in two (left column) and three (right column) dimensions in various confining boxes.
The wall densities are normalized to $1$, \ie, they have been divided by one minus the bulk fraction shown in Fig.~\ref{fig:bulk_fraction}.
In 2D, the boxes are ellipses with semi-axes $1$ and $0.1\le a\le10$. In 3D, they are ellipsoids with semi-axes $1$, $1$, and $0.1\le a\le10$.
As expected, in the persistent regime (top row, $\ell=100$)
the two analytical predictions agree with each other and with the data from each of the three models (ABP, AOUP, RTP).
It is also expected that Eq.~\eqref{eq:persistent_density} fails in the intermediate regime (bottom row, $\ell=1$), with the most prominent deviations observed in the regions of the boundary where the curvature is weakest, \ie, where the ratio of the persistence length to the local radius of curvature is the smallest (on the left of each graph). 
It is striking, on the other hand, that the good agreement between the three models and Eq.~\eqref{eq:aoup_density_stiff} survives in the intermediate regime.
From a practical point of view, it suggests that Eq.~\eqref{eq:aoup_density_stiff}, which has only been derived for AOUP, can be used to study the response to confinement over a wide range of persistence times in a larger class of self-propelled particles that includes ABP and RTP.
From a fundamental point of view, it means that this response can be understood through an equilibrium mapping given by~\eqref{eq:free_energy1} with $E=-\log\Omega$.

To complete our analysis of the density profile we plot in Fig.~\ref{fig:bulk_fraction} the fraction of particles in the bulk as a function of the persistence length in the simulations of Fig.~\ref{fig:radius-density}.
In agreement with Eq.~\eqref{eq:persistent_density} and Eq.~\eqref{eq:aoup_density_stiff}, the bulk fraction vanishes in the persistent limit for all three active models.
However, the way it decays toward zero is model-dependent. In ABP, it is roughly exponential. In RTP, it is algebraic with exponent $-1$. In AOUP, it is algebraic with exponent $-\D/3$. In RTP and AOUP, the effect of the box's aspect ratio can be captured by a single length scale $L$ (see below), as evidenced by the collapse of the simulation data in the corresponding panels when plotting against $\ell/L$. In ABP, we could not find such a single length scale and used $L=1$ instead. 

To make sense of the exponential decay in ABP, we return to the persistent theory of section~\ref{convex_abp} and quantify the deviations from $\vec{n}=\vec{u}$. The angle $\phi$ between $\vec{n}$ and $\vec{u}$ evolves according to~\footnote{The 2D version of Eq.~\ref{eq:abp_angle} was derived and analyzed in Ref.~\cite{Fily2014a}. The 3D version follows from the equation of motion for the tangential components of $\vec{\fa}$ derived in Ref.~\cite{Fily2016}.}
\begin{align}
\ddt{\phi} = -v \kappa \sin\phi + \frac{1}{\sqrt{(\D-1)\tau}} \xi(t)
\label{eq:abp_angle}
\end{align}
where $\kappa$ is the curvature of the wall along the particle's trajectory and $\xi$ is a white Gaussian noise with zero mean and $\langle\xi(t)\xi(t')\rangle=2\delta(t-t')$.
The first term on the right-hand side corresponds to the rotation of $\vec{n}$ as the particle glides along the wall. For a convex wall, $\kappa>0$, $\vec{n}$ rotates toward $\vec{u}$, and $\phi$ relaxes to zero as the particle approaches its target location.
The second term corresponds to the angular noise.
Leaving the wall requires $|\phi|>\pi/2$. If $\kappa$ is constant, Eq.~\eqref{eq:abp_angle} can be interpreted as an overdamped Kramers escape problem from the energy well $E/\kb T=-\alpha\cos\phi$ where $\alpha=(\D-1)v\tau\kappa$. In the persistent regime, the well is very deep and the escape rate (thus the bulk fraction) is dominated by $e^{-\alpha}$~\footnote{There are additional power law terms, whose form is a bit different from the usual Kramer's escape problem because the escape does not happen near a smooth maximum of the $E$, but they are not relevant at this level of accuracy.}.

For an RTP, the dynamics between two tumbles is that of an infinitely persistent ABP, but tumbles can leave the particle with any value of $\phi$, sending it to the bulk on average every other time (whenever $|\phi|>\pi/2$). In the persistent regime, crossing the bulk happens in one straight motion and takes a typical time $L/v$ where $L$ is the size of the box. Since tumbles happen on average every $\tau$, the relative time spent in the bulk scales as $\ell^{-1}=(v\tau)^{-1}$. If $a=1$ (circular or spherical box), the average length of a straight bulk crossing is simply the average chord length ($4/\pi$ for a circle, $1$ for a sphere) and the asymptote of the bulk fraction can be obtained exactly: $2/(\pi\ell)$ in 2D, $1/(2\ell)$ in 3D. Interestingly, rescaling $\ell$ by $L=2/\pi$ in 2D and $L=1/2$ in 3D collapses the 2D and 3D data onto each other for all values of the aspect ratio $a$. Furthermore, we find that multiplying the normalizing length scale $L$ by $a_+^{1/4}a_-^{3/4}$ where $a_+$ is the long semi-axis and $a_-$ is the short semi-axis collapses all available RTP data onto a single curve, as shown in Fig.~\ref{fig:bulk_fraction}. 

In the AOUP case, Eq.~\eqref{eq:aoup_density_stiff} provides a means to evaluate the bulk fraction. In the persistent regime, the integral of $\Omega$ is dominated by the boundary term proportional to $\ell^\D$: $\int_{\Bb}\kappa\ell^2/2=\pi\ell^2$ in 2D and $\int_{\Bb}\sqrt{\pi/54}K\ell^3=(2\pi/3)^{2/3}\ell^3$ in 3D. The integral over the bulk is simply the $\D$-dimensional volume of the box $V=\int_\B 1$. Thus we expect the bulk fraction to scale like $(\ell/L)^\D$ where $L=V^{1/\D}$.
In the ellipses and ellipsoids of Fig.~\ref{fig:bulk_fraction} this becomes $L=a^{1/\D}$, which is indeed the only relevant length scale as evidenced by the full collapse of the data onto a 2D and a 3D master curve. The predicted decay exponent, on the other hand, is three times too large.
To understand the decay exponent, we note that the magnitude $f'$ of $\vec{\fa}$ obeys $\ddt{f}' = -f'/\tau+f\xi(t)/\sqrt{\tau}$ with $\langle\xi(t)\xi(t')\rangle=2\delta(t-t')$ while its orientation $\vec{u}$ obeys Eq.~\eqref{eq:eom_fa} with an effective persistence time $\tau'=\tau (f'/f)^2$. Thus in the persistent limit, we may think of an AOUP as an ABP whose active force magnitude and persistence time change quasistatically. According to our ABP analysis, the probability of being in the bulk is controlled by the effective persistence length $\ell'=\mu f'\tau'=(f'/f)^3\ell$. Since $\vec{\fa}$ is Gaussian, the distribution of $f'$ is Maxwellian and that of $\ell'$ is
$$
p(\ell') =
\frac{2}{3\ell'}  \left(\frac{\ell'}{\ell}\right)^{\D/3}
\frac{\left(\dfrac{\D}{2}\right)^{\D/2}}{\Gamma\left(\dfrac{\D}{2}\right)}
\exp\left[ -\dfrac{\D}{2} \left(\dfrac{\ell'}{\ell}\right)^{2/3} \right]
$$
Using the ABP estimate $e^{-\ell'}$ for the probability of being in the bulk when the effective persistence length is $\ell'$ yields to the bulk fraction estimate $\int d\ell' p(\ell')e^{-\ell'}$. For large $\ell$ this expression scales like like $\ell^{-\D/3}$, in agreement with Fig.~\ref{fig:bulk_fraction}~\footnote{The decay exponent is not sensitive to the exact expression of the bulk fraction for a given $\ell'$. For example, the probability of $\ell'$ being smaller than any given length $L$ also scales like $\ell^{-\D/3}$ for large $\ell$.}.


\section{Nonconvex walls}
\label{nonconvex}

\subsection{Hard-wall persistent theory}
\label{nonconvex_abp}

In concave regions of the boundary ($\kappa<0$ in Eq.~\eqref{eq:abp_angle}; point $Q$ in Fig.~\ref{fig:wall_sketch}), the gliding motion of the particle along the wall causes $\vec{n}$ to rotate away from $\vec{u}$, making $\vec{n}=\vec{u}$ (ie, $\phi=0$) unstable. In the persistent limit, particles move quasistatically from stable equilibrium point to stable equilibrium point. Therefore, concave regions are empty. Furthermore, nonconvex boxes have multiple convex locations with the same normal vector. In section~\ref{convex_abp} we obtained the density profile by noting that, in the persistent limit, the location on the boundary with normal vector $\vec{n}$ inherits all the particles with orientation $\vec{u}=\vec{n}$. If there are multiple locations with the same normal, each location inherits only a fraction of those particles. 
This has two important consequences. First, the density does not depend solely on the local geometry any more; in particular, it depends on the existence and the number of other (distant) locations with the same normal vector as the point of interest.
This rules out a free energy of the form~\eqref{eq:free_energy1} with a local effective potential $\Ue$. Second, the density profile is no longer entirely determined by $\vec{u}=\vec{n}$. When a particle has multiple target locations, which one it goes to depends on its initial position, which itself depends on the history of its orientation, thus introducing a dependence on the dynamics.
In other words, nonconvexity undermines both the existence of an effective local free energy and the insensitivity to the details of the dynamics that characterize equilibrium systems.

\begin{figure}
\centering
\includegraphics[width=0.99\linewidth]{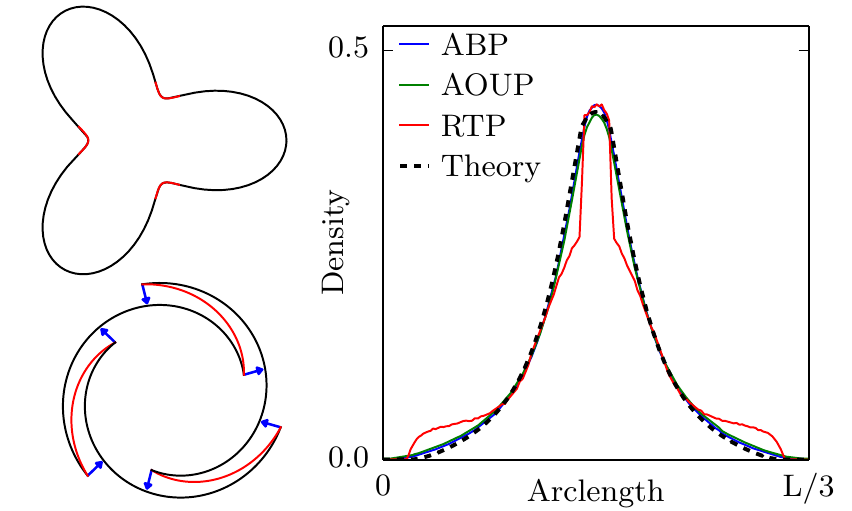}
\caption{
Density profile in a 2D nonconvex box deep in the persistent limit ($\ell=10^3$).
Top left: Shape of the box in real space. Convex (resp. concave) regions are shown in black (resp. red).
Bottom left: The box in normal vector space. Convex (resp. concave) regions are shown in black (resp. red). The distance from the center only serves to distinguish the regions where they overlap. When a particle reaches the end of a convex region, it jumps to a point with the same normal on a different, overlapping, convex region (blue arrows).
Right: Density on the boundary as a function of arclength in one of the box's three lobes. The dashed line is the theoretical prediction for ABP~\cite{Fily2015}. 
}
\label{fig:nonconvex}
\end{figure}

Let us review the case of a 2D ABP, which has been studied in depth~\cite{Fily2015}. As in section~\ref{convex_abp}, the problem is best visualized in the space of normal vectors.
In nonconvex boxes, however, we must ``unfold'' this space to distinguish between different physical locations with the same normal vector. This is illustrated in Fig.~\ref{fig:nonconvex} for the box with polar representation $r(\theta)=1+0.5\cos(3\theta)$. Convex (resp. concave) regions are shown in black (resp. red). Each concave region induces an overlap, \ie, a range of normal vectors that exist in the concave region and in both its neighboring convex regions. The dynamics within each convex region is the same as that of the orientation, \ie, free diffusion on the unit circle. When a particle reaches the end of a convex region (\ie, an inflexion point), however, it skips over the (unstable) concave region to the next convex location with the same normal (blue arrow in the bottom left panel of Fig.~\ref{fig:nonconvex}). A similar situation can be seen in Fig.~\ref{fig:wall_sketch}: a noiseless particle at $C$ whose orientation is ever so sightly rotated clockwise compared with the normal at $C$ will only come to a rest near the next convex location whose normal matches the one at $C$, \ie, near $A$. In the persistent limit, such jumps are instantaneous (compared with the diffusive dynamics in convex regions) and the density profile is obtained by solving a free diffusion equation in each convex region with a sink at each inflexion point and a corresponding source at the next location that has the same normal as the inflexion point~\cite{Fily2015}.

Although the RTP case has not been solved, it is clear that the structure of the master equation to solve to obtain the density is quite different from the one discussed above.
In particular, nonlocal couplings are not limited to a discrete set of jumps (one per inflexion point in the ABP case); rather, every point is coupled to nearly every other point as the orientation of an RTP can jump from any value to any other value at any time.


\subsection{Simulations}
\label{nonconvex_simulations}

Fig.~\ref{fig:nonconvex} shows the density profile on the boundary measured in Brownian dynamics simulations of ABP, RTP, and AOUP in a nonconvex box with polar representation $r(\theta)=1+0.5\cos(3\theta)$ deep in the persistent regime ($\ell=10^3$), along with the persistent limit ABP prediction from Ref.~\cite{Fily2015}. Both the ABP and the AOUP data agree with the ABP prediction, consistent with Ref.~\cite{Fily2015} and with the idea that a very persistent AOUP is analogous to an ABP whose self-propulsion speed and persistence time evolve quasistatically (see section~\ref{convex_simulations}). The RTP data, on the other hand, is qualitatively different, as suggested in section~\ref{nonconvex_abp}. In particular, it exhibits discontinuities at locations where the normal multiplicity (the number of convex locations having the same normal vector as the current point) changes.


\subsection{AOUP theory}
\label{nonconvex_aoup}

\newcommand{\M}{M}
\newcommand{\dbk}{\db_\kappa}
\newcommand{\dbp}{\db_p}
\newcommand{\km}{\kappa_\text{min}}

The nonlocality of the density in nonconvex boxes discussed above and in Ref.~\cite{Fily2015} is in direct contradiction with Eq.~\eqref{eq:aoup_density_general}. To resolve this apparent paradox, we now look at the implications of the positive definite condition on $\M_{ij}\equiv(\delta_{ij} + \tau \mu \partial_{ij} \U)$ that is needed to derive Eq.~\eqref{eq:aoup_density_general}~\cite{Maggi2015}. As in section~\ref{convex_aoup}, we start with the elastic wall potential defined in section~\ref{model} before taking the stiff limit.

In the bulk $\M_{ij}=\delta_{ij}$, whose only eigenvalue is $1$. Beyond the boundary, $\M_{ij} = \delta_{ij} + \lambda(n_i n_j + \db L_{ij})$  has eigenvalues $1+\lambda\tau$ and $\{1+\lambda\tau\db\kappa_a\}_{1\le a<\D}$. Thus $\M_{ij}$ is positive definite 1) everywhere in convex boxes, whose principal curvatures $\kappa_a$ are all positive everywhere, but 2) only within a distance $\dbk\equiv(\lambda\tau\km)^{-1}$ of the boundary in nonconvex boxes, where $\km = -\min\{\kappa_a\}_{1\le a<\D}$ is the absolute value of the most negative (most concave) principal curvature. Therefore, $M_{ij}$ is positive definite everywhere if and only if the box is convex. In other words, the theory breaks down in nonconvex boxes.

On the other hand, the density in the wall drops quickly beyond the penetration length $\dbp=\sqrt{D/(\lambda(1+\lambda\tau))}$ implied by the exponential term in Eq.~\eqref{eq:aoup_density_elastic}. If $\dbk\gg\dbp$, the region where $\M_{ij}$ has negative eigenvalues is inaccessible and should be irrelevant. For a stiff wall, this happens when $\ell\ll\km^{-1}$. Therefore, we expect Eq.~\eqref{eq:aoup_density_elastic} to remain applicable to nonconvex boxes in the quasi-thermal limit, \eg, by setting $\Omega=0$ beyond $\dbk$ where Eq.~\eqref{eq:aoup_density_general} suggests it would be (slightly) negative. In contrast, in the persistent limit the theory truly fails to capture the density profile, as evidenced by the discussion of 2D ABP in section~\ref{nonconvex_abp} and the agreement between the ABP and the AOUP data in Fig.~\ref{fig:nonconvex}.

Finally, it is instructive to look at the reason why Eq.~\eqref{eq:aoup_density_general} breaks down when $M_{ij}$ is not positive definite. The dynamics of an AOUP with position $\vec{x}$ in an arbitrary external potential $\U$ obeys~\cite{Maggi2015} 
\begin{align}
\tau\ddot{x}_i + M_{ij} \dot{x}_j = -\mu \partial_i \U + \sqrt{D} \xi_i(t)
\label{eq:aoup_langevin}
\end{align}
where $\xi_i$ is a white Gaussian noise with zero mean and $\langle\xi_i(t)\xi_j(t')\rangle=2\delta_{ij}\delta(t-t')$. The unified colored noise approximation used in Ref.~\cite{Maggi2015} neglects $\ddot{x}_i$, which leads to Eq.~\eqref{eq:aoup_density_general}. Interpreting $\tau$ as a mass and $M_{ij}$ as a friction coefficient, it is analogous to an overdamped approximation. In concave regions, the effective friction $M_{ij}$ is negative along the boundary. This invalidates the overdamped assumption, hence Eq.~\eqref{eq:aoup_density_general}. What is more, negative friction threatens the very idea of an equilibrium mapping.

Interestingly, the dynamics of an ABP along the wall of a nonconvex box can also be interpreted as that of a massive particle with a position-dependent effective friction that is negative in concave regions~\cite{Fily2015}. In fact, this negative friction can be seen as the reason why concave regions are unstable. It can also explain why particles jumping over concave regions only stop deep inside the next convex region: they accelerate over the entire concave region, only beginning to slow when they re-enter a convex region. In the persistent limit, when the friction is very large (in absolute value), this results in the nonlocal jumps between locations with the same normal discussed in section~\ref{nonconvex_abp} (\eg, from $C$ to $A$ in Fig.~\ref{fig:wall_sketch}).


\section{Conclusion}
\label{conclusion}

Polar-isotropic active matter has attracted major attention in recent years as a testing ground for the application of equilibrium mapping techniques to active systems. Here we have discussed the existence of such equilibrium mappings in three models of noninteracting polar-isotropic active particles confined within hard walls, by looking at two hallmarks of equilibrium behavior: 1) the independence of steady-state on the detailed particle dynamics and 2) the fact that the density profile in noninteracting systems minimizes a free energy of the form~\eqref{eq:free_energy1}.

In convex confining boxes, both properties are verified exactly in both the quasi-thermal limit (zero persistence time) and the persistent limit (infinite persistence time), and approximately in the intermediate regime. In the latter, the three models we studied exhibit an emptying of the bulk and an accumulation of particles in regions where the wall is most curved as the persistence time is increased. The affinity for curved regions is mostly model-independent and well captured by the AOUP theory of Ref.~\cite{Maggi2015}, which in the persistent limit is equivalent to the ABP theory of Refs.~\cite{Fily2014a,Fily2016}. In particular, those theories allow to derive explicitly the effective external potential $\Ue$ in Eq.~\eqref{eq:free_energy1}. The emptying of the bulk, on the other hand, depends on the model: in ABP the population of the bulk decays exponentially with the persistence time whereas in RTP and AOUP the decay is algebraic, consistent with previous results on ABP and RTP in 2D isotropic harmonic traps~\cite{Solon2015b}. This is most relevant in the crossover regime, when the persistence length is comparable to the size of the box and the bulk is neither nearly full nor nearly empty.

In nonconvex boxes, both the equivalence between models and the existence of a local effective potential are lost in the persistent regime. The density profile on the boundary becomes nonlocal and develops discontinuities for RTP but not ABP or AOUP. This change of behavior is accompanied by a change in the mathematical structure of the main theories used to describe polar-isotropic active particles. The ABP theory is concerned with the dynamics along the boundary, which it recasts as a thermal-like dynamics with a geometry-dependent friction that becomes negative in concave regions of the boundary~\cite{Fily2015}. Similarly, the breakdown of the AOUP theory traces back to the effective friction coefficient becoming negative in the (exact) equation of motion for the position when the potential is nonconvex and the persistence time is large [Eq.~\eqref{eq:aoup_langevin}]. This suggests that the method used to study a persistent ABP in a nonconvex box may be applicable to an AOUP in smooth nonconvex external potentials. Most importantly, it suggests a profound connection between the convexity of the external potential and the existence of equilibrium mappings in persistent active particles.

Understanding this connection could have practical as well as fundamental implications, since large persistence and nonconvex boundaries are the two key ingredients of many applications involving self-propelled particles, including their trapping by wedges~\cite{Kaiser2012,Kaiser2013,Guidobaldi2014} or asymmetric barriers~\cite{Koumakis2013}, their directed motion along a channel~\cite{Galajda2007,Wan2008,Tailleur2009,Nikola2016}, and their use to power micro-motors~\cite{Angelani2009,DiLeonardo2010,Sokolov2010} or propel passive objects~\cite{Mallory2014a}. 


\begin{acknowledgments}
We thank Julien Tailleur for discussions. We acknowledge support from the NSF, DMR-1149266, the Brandeis Center for Bioinspired Soft Materials, an NSF MRSEC, DMR-1420382, and from the W. M. Keck Foundation. Computational resources were provided by the NSF through XSEDE computing resources and the Brandeis HPCC.
\end{acknowledgments}


\bibliography{abp_vs_gcn}

\end{document}